\documentclass{aastex631}
\usepackage{hyperref} 
\usepackage{natbib}
\usepackage{textcomp}
\usepackage{gensymb}
\usepackage{xcolor}
\usepackage{amsmath}
\usepackage{comment}
\usepackage{graphics}
\usepackage{array}
\usepackage{tabularx}
\usepackage{booktabs}
\usepackage{amsmath,amssymb}
\usepackage{tablefootnote}
\usepackage[utf8]{inputenc}
\usepackage{newunicodechar}

\newunicodechar{ʼ}{'}

\begin{document}
\title{The Milky Way is a less massive galaxy---new estimates of the Milky Way's local and global stellar masses}
\author[0000-0001-5258-1466]{Jianhui Lian}
\affiliation{South-Western Institute for Astronomy Research, Yunnan University, Kunming, Yunnan 650091, People’s Republic of China}
\affiliation{Key Laboratory of Survey Science of Yunnan Province, Yunnan University, Kunming, Yunnan 650500, Peopleʼs Republic of China}
\email{jianhui.lian@ynu.edu.cn (JL)}  

\author{Tao Wang}
\affiliation{South-Western Institute for Astronomy Research, Yunnan University, Kunming, Yunnan 650091, People’s Republic of China}
\affiliation{Key Laboratory of Survey Science of Yunnan Province, Yunnan University, Kunming, Yunnan 650500, Peopleʼs Republic of China}

\author{Qikang Feng}
\affiliation{Department of Astronomy, School of Physics, Peking University,
 Beijing 100871, People’s Republic of China}
\affiliation{Kavli Institute for Astronomy and Astrophysics, Peking University,
 Beijing 100871, People’s Republic of China}

\author{Yang Huang}
\affiliation{School of Astronomy and Space Science, University of Chinese
 Academy of Sciences, Beijing 100049, People’s Republic of China}
\affiliation{National Astronomical Observatories, Chinese Academy of Sciences, Beijing 100012, People’s Republic of China}

\author{Helong Guo}
\affiliation{South-Western Institute for Astronomy Research, Yunnan University, Kunming, Yunnan 650091, People’s Republic of China}
\affiliation{Key Laboratory of Survey Science of Yunnan Province, Yunnan University, Kunming, Yunnan 650500, Peopleʼs Republic of China}


\begin{abstract}
Stellar mass is the most fundamental property of a galaxy. While it has been robustly measured for millions of external galaxies, it remains poorly constrained for the Milky Way because of the strong selection effect from our inside perspective. In this work, we reconstruct the intrinsic vertical mass density profile in the solar neighborhood and the radial mass density profile across the entire Galaxy using data from the Gaia and APOGEE surveys, following careful correction for the selection function. The local density profile exhibits strong north-south asymmetry in the geometric thick disk regime and increases steeply toward the disk mid-plane, favoring an exponential model over the sech$^2$ model. Integrating the local vertical density profile yields a surface stellar mass density of 31.563$\pm$2.813(syst.)$\pm$0.024(stoch.)~${\rm M_{\odot}pc^{-2}}$, of which 25.074 and 6.489~${\rm M_{\odot}pc^{-2}}$  
correspond to living stars and stellar remnants, respectively.  
The radial surface mass density profile {of the Milky Way} shares the same flat inner component as observed in the brightness profile. With this mass density profile {and local mass density from Gaia}, we derive a new estimate of the total stellar mass of the Milky Way of { 2.607$\pm$0.353(syst.)$\pm$0.085(stoch.)${\rm \times10^{10}M_{\odot}}$}, a factor of two lower than the previous results. This discrepancy arises primarily from the inner disk profile, which was previously unavailable and extrapolated from the outer disk profile. 
The lower stellar mass estimate of the Milky Way significantly reduces its rarity in terms of supermassive black hole mass among external galaxies { and implies a larger dark matter-to-baryon mass ratio in the inner Galaxy}.  
\end{abstract}

\section{Introduction}
Stellar mass is the most fundamental property of galaxies. It has the broadest correlation among galaxy global properties, such as the mass-metallicity and mass-[Mg/Fe] relations \citep{tremonti2004,thomas2010}, the mass-star formation rate relation \citep{brinchmann2004}, the mass-size relation \citep{shen2003}, and the stellar mass-supermassive black hole mass connection \citep{reines2015}. Such relationships underscore the unparalleled importance of stellar mass in regulating and probing the evolution of galaxies. The prevalence of multi-band photometric and spectroscopic observations of galaxies, coupled with improvements in stellar population synthesis techniques, have enabled robust measurements of  stellar mass of external galaxies from the high-redshift to the local Universe.                              

Living inside the Milky Way gives us the ability to resolve individual stars and study detailed substructures \citep[e.g.,][]{shen2020,poggio2021} and the evolution history of the Milky Way \citep[e.g.,][]{chiappini1997,spitoni2019,lian2020a,sharma2020}. For the same reason, however, we lack an outside perspective of the Milky Way, making it challenging to measure the Galactic global properties. Our position in the disk results in highly biased and incomplete observations, favoring nearby stars and missing those towards the inner Galaxy, where the dust extinction is high and most stars are located.  

This situation has been rapidly changing thanks to the advent of massive stellar spectroscopic surveys (e.g., LAMOST, \citealt{zhao2012}; APOGEE, \citealt{majewski2017}; GALAH, \citealt{martell2016}) and Gaia astrometric and photometric survey. 
In the early stage of these massive surveys, when the sample size is limited, a forward modelling approach has been used to account for the selection function and understand the large-scale structure of the Milky Way \citep{bovy2012,bovy2016,mackereth2018,imig2022}. This method preassumes a global density model of the Milky Way, convolved with the survey selection function, to predict the observed spatial distribution of stars. This approach relies on the preassumed density model and the results are unavoidably model dependent. As the observed sample size increases rapidly, an alternative approach has also been proposed that directly divides the observed density of stars by the probability of being observed (i.e., essentially the selection function) to obtain the intrinsic density distribution of the underlying stellar populations \citep{xiang2018,yu2021,lian2022b}. 

The Galactic large-scale density distribution is a premise for estimating the Milky Way's global properties. Based on the 3D intrinsic density distribution reconstructed by \citet{lian2022b}, \citet{lian2023} derived the first measurement of the integrated average metallicity profile of the Milky Way, which exhibits a `$\wedge$'-like broken shape that is relatively uncommon in local galaxies. In a follow-up work, \citet{lian2024b} recovered the surface brightness profile of the Milky Way over an unprecedented radial range from R=0 to 17~kpc and obtained a half-light radius of 5.75$\pm$0.38~kpc, which is significantly larger than expectations from earlier scale length measurements, but rather consistent with local galaxies of similar mass. {Interestingly}, this surface brightness profile also presents a break around 7.5~kpc, with the inner disk profile being much flatter than the outer disk. 
In terms of stellar mass, \citet{licquia2015} estimated the stellar mass of the disk by assuming the single-exponential disk density model from \citet{bovy2013} and that of the bulge from merging the measurements in the literature using a Bayesian statistical method. They obtained a disk stellar mass of 5.17$\pm1.11{\rm\times10^{10}\ M_{\odot}}$, a bulge stellar mass of 0.91$\pm0.07{\rm\times10^{10}\ M_{\odot}}$, and {in total} 6.08$\pm1.14{\rm\times10^{10}\ M_{\odot}}$. \citet{xiang2018} delivered a moderately lower stellar mass of the disk of $\sim$4${\times10^{10}\ M_{\odot}}$ based on their surface density profile reconstructed between R=7 to 12~kpc using data from the LAMOST survey and extrapolation of the profile to the inner Galaxy at R$<7$~kpc. Compared to the global stellar mass, measuring the local stellar mass {surface} density in the solar neighborhood is less challenging. {Consistent measurements around} 33--36~${\rm M_{\odot}pc^{-2}}$, including visible stars and stellar remnants, have been reported in the literature \citep[][]{flynn2006,mckee2015,xiang2018}. 

In this work, we aim at improving the local and global mass estimate of the Milky Way by using the unprecedented data from Gaia and APOGEE surveys and applying a careful correction for the selection function. In particular, we will investigate the impact of the broken disk density profile on the measurement of the Milky Way's global mass. Unless otherwise stated, we adopt a solar position with Galactocentric radius of 8.2~kpc and height 0.027~kpc above disk plane \citep{bland2016}.  

\section{Data}
In this work, we use Gaia data to calibrate the local surface mass density and APOGEE data to reconstruct the radial surface mass density profile, and combine these two results to derive the total stellar mass of the Milky Way. 

\subsection{Gaia data}
For Gaia data, we select stars with robust distance indicated by low parallax error (parallax\_error/parallax$<$0.2) and located in a cylinder perpendicular to the Galactic disk plane and centered on the Sun within a cylinder radius of 0.3~kpc and height of 3~kpc. This cylinder is then separated into series of heights from -3 to 3~kpc, with a step of 0.05~kpc between [-3, -1.5] and [1.5, 3]~kpc and step of 0.01 kpc between [-1.5, 1.5]~kpc. This height binning scheme ensures dense enough sampling close to the disk plane and robust calculation of the observed {stellar} density at large height. With the parallax criteria, the observed sample is complete down to 15.5~mag in Gaia RP band throughout the cylinder region. To facilitate the calculation of selection function at short height bins, we apply a selection criterion of the sample in absolute magnitude in RP, instead of in apparent magnitude, as usually adopted in previous works. For each height bin, we calculate the absolute magnitude limit above which all stars within this bin should be observed given the apparent magnitude limit of 15.5~mag and 3D extinction map from \citet{guo2021} as:
\begin{equation}
    M_{\rm RP,limit} = 15.5-5{\rm log}(D_{\rm max}/0.01)-1.5\sigma_{\rm ext},
\end{equation}
where $D_{\rm max}$ and $\sigma_{\rm ext}$ represents the maximum distance and the standard deviation of extinction measurements from the 3D extinction map within this height bin, respectively. We do not use the maximum extinction to avoid stochastic extreme measurements which will dramatically reduce the number of selected stars. 
{Finally, 3 million stars are selected in our Gaia sample.}
The observed number density of stars at a given height bin is equal to the observed number of stars divided by the volume of this bin. 

To estimate the intrinsic density of stars, we first calculate the fraction of underlying stellar population that are observed and selected in our sample, i.e. essentially the selection function of the sample, and then divide the observed stellar density by this fraction. This selection function is estimated through sampling the theoretical stellar isochrones and applying the selection criteria adopted for the observed sample. As age and metallicity information is not available for most of the selected stars from Gaia, we are unable to estimate and correct for the selection function for individual mono-age or mono-abundance populations as in previous works using data from spectroscopic surveys \citep[e.g.,][]{bovy2016,xiang2018,lian2022b,imig2022}. Alternatively, we generate a single mock catalog of all stellar populations by sampling PARSEC stellar isochrones \citep{bressan2012} using the age-[Fe/H] distribution corrected for selection function \citep{lian2022b} and Kroupa initial mass function within 0.1 to 100~$\rm M_{\odot}$ \citep{kroupa2001}. Applying the absolute magnitude cut described in Eq.(1), we obtain the fraction of living stars that are in our selected sample. Dividing this observed number of stars by this fraction and times the mass-to-number ratio of the mock catalog, we finally derive the intrinsic mass density of alive stars at each height bin. We then also calculate the mass density of stellar remnants, including white dwarf, neutron stars, and black holes, using the mass ratios of the living star to the remnants as a function of age and metallicity from the stellar evolution model of  \citet{maraston2005}.  

\subsection {APOGEE data}
APOGEE survey, which operates in the near-infrared, is currently the only Galactic spectroscopic survey that delivers observations of stars over the full radial range of the Milky Way, from the Galactic center to the edge of the outer disk \citep{zasowski2013,zasowski2017,beaton2021,santana2021}. We take chemical abundance measurements from the APOGEE catalog \citep{nidever2015,garcia2016,smith2021} and distance and age measurements from the astroNN value-added catalog for APOGEE stars \citep{mackereth2019,leung2019} {from SDSS Data Release 17 \citep{blanton2017,sdss-dr17}}. Here we follow our previous work to derive the intrinsic density of the disk by directly correcting for the selection function \citep{lian2022b}. We briefly summarize the major steps here and refer the readers to {that} paper for more details. The whole process includes three main steps: {selecting an clean observed sample, calculating and then correcting forthe selection function}. With precise abundance information of individual stars from {high-qaulity spectra}, we perform these steps for mono-abundance populations (defined in [Mg/Fe] and [Fe/H]) separately to recover the density distribution of stars born at different epochs. We first select giant stars in the main survey  {following the criteria listed in \citet{lian2025a}} and bin them into a multi-dimension space in [Mg/Fe], [Fe/H], $l$, $b$, and distance. 
Accordingly, we calculate the selection function in the same multi-dimension space. The selection function is a convolution of three successive components, the fraction of stars falling within the APOGEE candidate selection box in the 2MASS color-magnitude diagram, the fraction within this box that is targeted, and the fraction of observed stars entering into our final selected sample. We calculate the first and third components by ourselves and take the calculation of the second component from \citet{imig2022}. 
The intrinsic density of stars is simply derived by dividing the observed number density by the obtained selection function in the multi-dimension space in [Mg/Fe], [Fe/H], $l$, $b$, and distance. The conversion from number density to mass density and calculation of the mass density of stellar remnants are the same as that described above for the Gaia data. 

\section{Results and Discussion}
\subsection{Local stellar mass}
\begin{figure*}
	\centering
	\includegraphics[width=\textwidth]{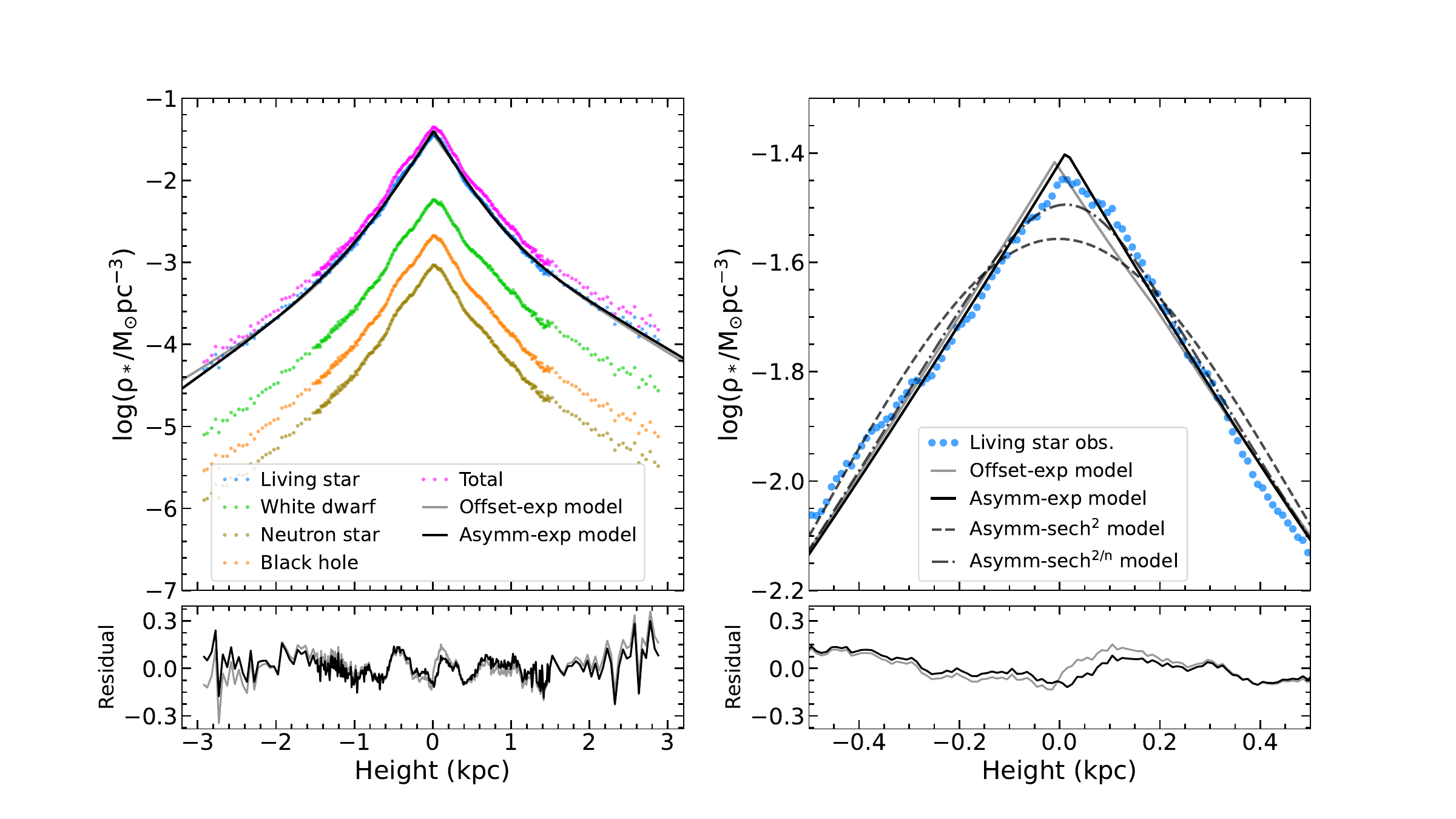}
	\caption{Vertical mass density profiles of living stars and stellar remnants. Asymmetric {and offset density models} with exponential, sech$^2$, and sech$^{2/n}$ profiles are applied to fit the {vertical density profiles} as shown in the right panel. In the left panel, only the best-fitted exponential model for the living stars is included as shown in black line.}
    \label{lmp}
\end{figure*}

\begin{table*}
    \caption{Best-fitted parameters of the local vertical density models.}
    \centering
    \begin{tabular}{l c c c c c c c c c }
    \hline
Asymm.  & n & $z_0$ & $\rho_{0,thin}$ & $h_{z,thin,n}$ & $h_{z,thin,s}$ & $\rho_{0,thick}$ & $h_{z,thick,n}$ & $h_{z,thick,s}$ & $\Sigma_*$  \\
 models  & & pc & ${\rm M_{\odot}/pc^3}$ &  kpc & kpc & ${\rm M_{\odot}/pc^3}$ & kpc & kpc & ${\rm M_{\odot}/pc^2}$  \\
 \hline
exp & $\infty$ & 13.3$\pm$1.9 & 0.036$\pm$0.0002 & 0.26$\pm$0.004 & 0.27$\pm$0.005 & 0.0042$\pm$0.0004 & 0.77$\pm$0.03 & 0.64$\pm$0.018 & 25.07$\pm$0.03 \\
${\rm sech^2}$ & 1 & -2.4$\pm$0.9 & 0.025$\pm$0.0001 & 0.19$\pm$0.002 & 0.18$\pm$0.002 & 0.0026$\pm$0.0001 & 0.61$\pm$0.011 & 0.53$\pm$0.007 & 24.57$\pm$0.02 \\
${\rm sech^{2/n}}$ & 3.6$\pm$0.1 & 12.9$\pm$1.5 & 0.102$\pm$0.004 & 0.07$\pm$0.001 & 0.07$\pm$0.001 & 0.0131$\pm$0.0005 & 0.2$\pm$0.0001 & 0.17$\pm$0.002 & 24.86$\pm$0.02 \\
    \hline
    \hline
Offset  & n & $z_{0,thin}$ & $\rho_{0,thin}$ & $h_{z,thin}$ & $z_{0,thick}$ & $\rho_{0,thick}$ & $h_{z,thick}$ & $\Sigma_*$  \\
 models  & & pc & pc & ${\rm M_{\odot}/pc^3}$ & kpc & ${\rm M_{\odot}/pc^3}$ & kpc & ${\rm M_{\odot}/pc^2}$  \\
 \hline
exp & $\infty$ & -9.5$\pm$1.2 & 0.036$\pm$0.0002 & 0.28$\pm$0.004 & 191.7$\pm$18.5 & 0.0033$\pm$0.0004 & 0.75$\pm$0.03 & 24.9$\pm$0.04 \\
${\rm sech^2}$ & 1 & -4.0$\pm$1.0 & 0.025$\pm$0.0001 & 0.19$\pm$0.002 & 133.6$\pm$7.0 & 0.0025$\pm$0.0001 & 0.57$\pm$0.008 & 24.52$\pm$0.02 \\
${\rm sech^{2/n}}$ & 3.5$\pm$0.1 & -9.9$\pm$1.2 & 0.1$\pm$0.0044 & 0.07$\pm$0.001 & 163.3$\pm$0.02 & 0.0102$\pm$0.0007 & 0.2$\pm$0.0003 & 24.76$\pm$0.05 \\
    \end{tabular}
    \label{t1}
\end{table*}
The derived intrinsic mass density of living stars and stellar remnants as a function of height at the solar radius is shown in Figure~\ref{lmp}. A clear north-south asymmetry is present in the vertical mass density profile with the density decreasing more slowly in the north ({i.e. positive height}). Unlike the asymmetry found in \citet{dobbie2020}, this asymmetry only appears at large height beyond 1~kpc. In addition to the large-scale asymmetry, there is small-scale wave pattern in the density profile, as reported in many previous works  \citep{widrow2012,yanny2013,bennett2019,wangtao2024}. Both the oscillations and the asymmetry features are likely the result of perturbations induced by infalling dwarf galaxy. Given the presence of the asymmetry feature, we fit asymmetric density models to the data with independent scale heights in the north and south directions \citep{arellano-valle2005-eco,zhu2009-eco,arnroth2023-eco}. {In addition, we also explore an alternative density model in which we remove the north-south asymmetry in scale height but set the offset of the thick disk's mid-plane independent from that of the thin disk. This model is referred to as the `offset model' in this paper.}

The high quality of the distance and large number of observed stars allow us to investigate the local mass density profile on a very small scale. We fit the {local vertical density profile with two sets of models, one with different scale heights in the north and south and the other with independent offsets between the geometric thin and thick disks. Each set of models includes three different forms, including an exponential model, sech$^2$ model, and sech$^{2/n}$ model.} The former two models are specific cases of the sech$^{2/n}$ model when $n=\infty$ and 1, respectively. 
The {asymmetric} sech$^{2/n}$ model can be written as:
\begin{equation}
    \rho(z) = 
    \begin{cases}
    \frac{\rho_{\rm 0,thin}}{n\times cosh(\frac{|z-z_0|}{2h_{\rm z,thin,n}})^{2/n}}+\frac{\rho_{\rm 0,thick}}{n\times cosh(\frac{|z-z_0|}{2h_{\rm z,thick,n}})^{2/n}} & z>=z_0, \\
    \frac{\rho_{\rm 0,thin}}{n\times cosh(\frac{|z-z_0|}{2h_{\rm z,thin,s}})^{2/n}}+\frac{\rho_{\rm 0,thick}}{n\times cosh(\frac{|z-z_0|}{2h_{\rm z,thick,s}})^{2/n}} & z<z_0,
    \end{cases}
\end{equation}
where $\rho_{\rm 0}$ and $h_{\rm z}$ represents the density on the mid-plane ($z=z_0$) and the scale height, respectively. The best-fitted parameters and the integral surface mass density of each model are listed in Table~\ref{t1}. The best-fitted {asymmetric and offset exponential} models are also included in the right panel of Figure~\ref{lmp}. Consistent with visual inspection, the thick disk exhibits a significant asymmetry with the north part being thicker, whereas the north-south difference in the thin disk is negligible. {Interestingly, the offset models exhibit comparable goodness in fitting the vertical density profile. To disentangle these two possibilities requires additional information (e.g., chemical abundances) to identify the geometric thick disk stars embedded in the thin disk regime. This is beyond the scope of this work, but interesting to explore in the near future.} 

The observed local density profile imposes strong constraints on the form of the vertical density profile of the disk. As illustrated in Figure~\ref{lmp}, the sech$^2$ model is strongly disfavored by the data with a profile that is too smooth compared to the mass densities around the mid-plane. In contrast, the exponential model provides a reasonably good fit to the density profile in both small scale around the mid-plane and large scale up to a few kpc. When setting the index $n$ free, the best-fitted sech$^{2/n}$ model also favors a value much larger than one. To be consistent with the density models beyond the solar neighborhood, we adopt the asymmetric exponential model to estimate the local surface mass density. In addition to living stars, we obtain a surface mass density of 4.31, 0.66, 1.51~${\rm M_{\odot}/pc^2}$ for the white dwarf, neutron star, and black hole, respectively. In total, the local stellar surface mass density is 31.563$\pm2.813$(syst.)$\pm$0.03(stoch.)~${\rm M_{\odot}/pc^2}$.  

We follow the approach described in \citet{lian2024b} to estimate the stochastic and systematic uncertainties in the measurement of the stellar mass density. The stochastic uncertainty originates from the uncertainties in the observed mass density at each 3D position, which is assumed to be Poisson error. The uncertainties in Table~\ref{t1} only include the stochastic uncertainties of the measurement. Compared to stochastic uncertainties, systematic uncertainties are more difficult to estimate, {which in this case} come mainly from the calculation of the selection function, in particular the choice of the stellar evolution model and the 3D extinction map. Since the extinction is generally low in the solar neighborhood, here we only consider the systematics introduced by the choice of stellar evolution model. Following \citet{lian2024b}, we calculate the differences in the measurements when adopting a different stellar evolution model and assume the difference as a rough estimate of the systematic uncertainty. Using MESA isochrones \citep{choi2016,dotter2016}, we obtain a surface mass density 2.81 ${\rm M_{\odot}/pc^2}$ lower than the current result. 

The local mass density measurements derived in this work are in agreement with previous results with a slightly lower estimate and a much higher precision. \citet{mckee2015} found a total local surface density of stars and stellar remnants of 33.4$\pm$3~${\rm M_{\odot}pc^{-2}}$, of which 27.0$\pm2.7\ {\rm M_{\odot}pc^{-2}}$ is in visible stars and 4.9$\pm0.6\ {\rm M_{\odot}pc^{-2}}$ in white dwarfs. A slightly higher estimate of $\sim36$~${\rm M_{\odot}pc^{-2}}$ was reported in \citet{flynn2006} and \citet{xiang2018}. Regarding the density of visible stars alone, diverse results have been reported in the literature, ranging from 20 to 32~${\rm M_{\odot}pc^{-2}}$ \citep{bovy2013,mackereth2017,lian2022b}.  

\subsection{Global stellar mass}
\begin{figure*}
	\centering
	\includegraphics[width=\textwidth]{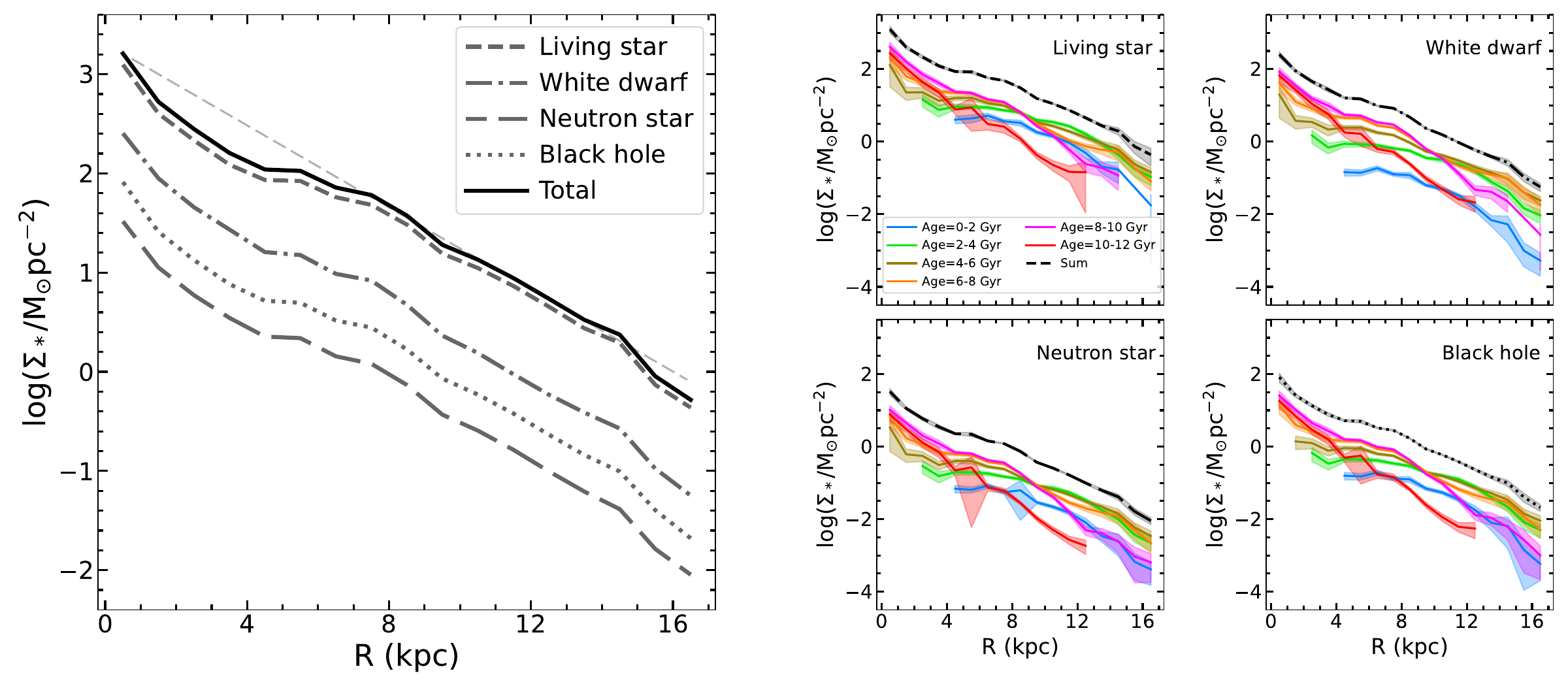}
	\caption{Radial surface mass density profile of various stellar objects. The { thin dashed} line in the left panel represents an exponential profile with scale length of 2.1~kpc. Each type of stellar objects is separated into mono-age populations shown in colorful lines in the right four panels. Shade regions indicate 1$\sigma$ uncertainties of the density measurements. }
    \label{gbp}
\end{figure*}

\begin{figure*}
	\centering
	\includegraphics[width=14cm]{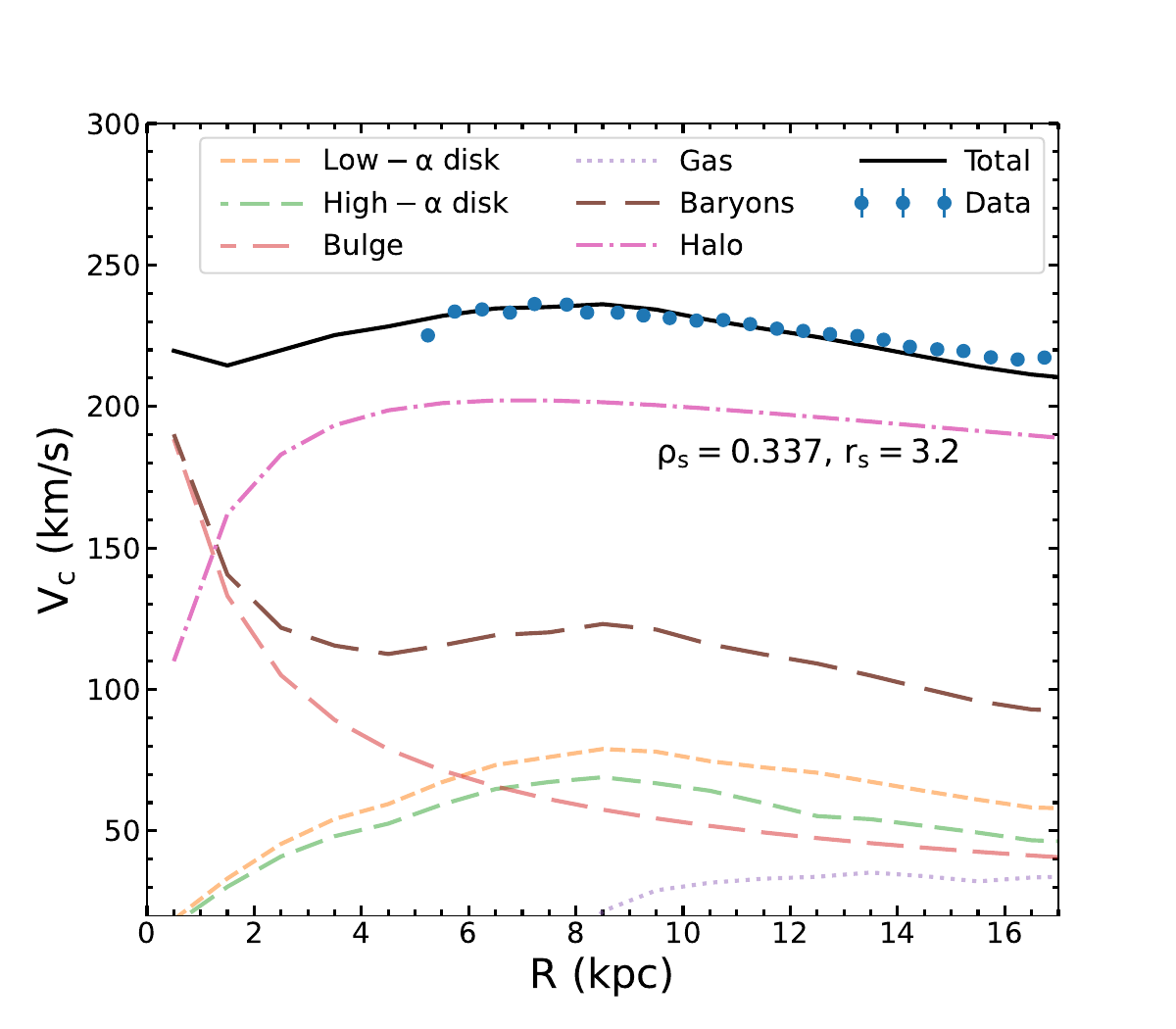}
        \caption{ Circular velocity decomposition based on our new stellar mass surface density profile. Observational results are taken from \citet{zhou2023}. The disk component is calculated based on the mass profile presented in Figure~\ref{gbp}. We separate the disk into low- and high-$\alpha$ parts based on a chemical definition. The bulge and gas disk components are calculated based on the density profiles taken from \citet{binney2011}. 
        The sum of all baryon components are denoted as the longest dashed curve in brown. The contribution of dark matter is set free and fitted assuming a NFW mass density profile with the best-fitted scaling radius ($r_s$) and characteristic density ($\rho_s$) presented in the figure.}
    \label{vc}
\end{figure*}

\begin{table*}
    \caption{Half-mass radii and integral stellar masses of different stellar remnants in the Milky Way.}
    \centering
    \begin{tabular}{l c c c c c c}
    \hline
Measurement & Unit & Living star & White dwarf & Neutron star & Black hole & In total \\
Half-mass radius & kpc & 4.25$\pm$0.19 & 3.62$\pm$0.16 & 3.97$\pm$0.19 & 3.96$\pm$0.2 & 4.12$\pm$0.15 \\
Integral stellar mass & ${\rm M_{\odot}}$ & 2.012$\pm$0.106 & 0.412$\pm$0.019 & 0.056$\pm$0.003 & 0.13$\pm$0.006 & 2.607$\pm$0.11 \\
    \hline 
    \end{tabular}
    \label{t2}
\end{table*}

In a companion paper, we present the radial surface brightness profile of mono-age populations of the Milky Way derived following the same procedure as in this work for the surface mass density profile \citep{lian2024b}. In that work, we found that the radial density profile of the Milky Way deviates from a single exponential profile with a flattened inner disk component. As a result, the half-light radius measured from the new radial density profile is significantly larger than that estimated based on the single-exponential disk model. 

Here we demonstrate that another critical implication of the flat inner disk profile is the overestimate of the total stellar mass of our Galaxy. {We first derive} surface mass density profiles of living stars and stellar remnants {as} presented in Figure~\ref{gbp}. For living stars, the mass density profile is qualitatively the same as the luminosity density. The outer disk between 8 and 14~kpc can be well described by a single exponential profile with a scale length of { 2.1}~kpc, which is shorter than the luminosity density profile of 2.6~kpc \citep{bland2016} because of negative gradient of age and thus mass-to-light ratio in the disk. The inner disk profile within solar radius also flattens, deviating from the exponential profile extrapolated from the outer disk. Again, because of the negative age gradient, the {inner flat feature is less prominent} in the case of {stellar mass than the luminosity profile}. For other remnants, the surface mass density profile and its age dependence generally follow that of living stars. The white dwarfs' mass density profile exhibits the strongest age dependence because their progenitors are low mass stars with a wide range of lifespan covering different age bins. Therefore, the fraction of stars {evolved} into white dwarfs decreases with age. 

The local surface mass density at the solar radius of all stellar remnants derived from APOGEE data is { 43.218~${\rm M_{\odot}/pc^2}$, significantly larger} than that of 31.563~${\rm M_{\odot}/pc^2}$ obtained using Gaia data. { The reason for the discrepancy is still unknown}. Given the substantially larger sample and the more precise distance from Gaia compared to the APOGEE data, we believe that the measurement with Gaia is more accurate and precise. Therefore, we recalibrate the surface mass density profile obtained with APOGEE data {by} the local measurement {in the solar neighborhood} with Gaia.  

Based on the recalibrated surface mass density profiles, we obtain the half-mass radius and integral stellar mass of living stars and stellar remnants. The results are included in Table~\ref{t2}. The half-mass radius of living stars is shorter than their half-light radius ({ 4.12} vs. 5.75~kpc) \citep{lian2024b} because of the negative age gradient. White dwarfs have the shortest half-mass radius among all remnants since a larger fraction of them originates from older populations, which have a {more} compact radial distribution. 

We obtain a total stellar mass of the Milky Way of { 2.607$\pm$0.353(syst.)$\pm$0.148(stoch.)${\rm \times10^{10}\ M_{\odot}}$}, of which { 2.012, 0.412, 0.056, and 0.13~${\rm \times10^{10}\ M_{\odot}}$} are that of living stars, white dwarfs, neutron stars, and black holes, respectively. When decomposing into disk and bulge, their stellar masses are { 1.935 and 0.673~${\rm \times10^{10}\ M_{\odot}}$}, respectively, implying a bulge-to-total ratio (B/T) of { 0.26}. The bulge mass is estimated as the sum of the surface mass density profile up to { $R=4$~kpc} after subtracting the disk contribution that is extrapolated from a linear fit to the disk profile between { 4} and 7~kpc.  

The stochastic and systematic uncertainties of the half-mass radius and the integral stellar mass are estimated as the local surface mass density above. The stochastic uncertainty originates from the uncertainties in the observed mass density at each {3D} position, the {radial position of} each radial bin, and the recalibration. The first part is assumed to be Possion error. The second part is assumed to be the standard deviation of the radii of individual 3D mass density measurements within each radial bin. The third part is taken from the local surface mass density measurement. Monte Carlo simulation with 100 trials is implemented to derive the effective stochastic uncertainty. 
Table~\ref{t2} only includes the stochastic uncertainties of the measurement. Regarding the systematic uncertainties, following \citet{lian2024b}, we calculate the differences in the measurements when adopting a different stellar evolution model and 3D extinction map in the bulge (where the extinction is most uncertain) and assume the difference as a rough estimate of the systematic uncertainty. Using MESA isochrones \citep{choi2016,dotter2016}, we obtain a half-mass radius 0.18~kpc larger and integral stellar mass 0.329${\rm \times10^{10} M_{\odot}}$ lower than the current results. Adopting the extinction map of the bulge from \citet{chen2013} results in a half-mass radius  0.35~kpc shorter and integral stellar mass 0.129${\rm \times10^{10}\ M_{\odot}}$ higher. Thus, we assume a systematic uncertainty of 0.39~kpc (i.e., $\sqrt{0.18^2+0.35^2}$) and 0.353${\rm \times10^{10} M_{\odot}}$ (i.e., $\sqrt{0.329^2+0.129^2}$) in our measurement of the half-mass radius and integral stellar mass of the entire stellar content. 

Our stellar mass estimates are {lower than the previous results by a factor of two}. {The inconsistency persists even considering the higher local mass density estimated with APOGEE data, which suggests a total stellar mass of the Milky Way of 3.55 ${\rm \times10^{10} M_{\odot}}$.} By analyzing the measurements from the literature using a Bayesian statistic approach, \citet{licquia2015} obtained a total stellar mass of 6.08$\pm$1.14${\rm \times10^{10} M_{\odot}}$ for the Milky Way, of which 0.91$\pm$0.07${\rm \times10^{10}\ M_{\odot}}$ from the bulge and 5.17$\pm$1.11${\rm \times10^{10}\ M_{\odot}}$ from the disk. For the bulge, the stellar mass constrained by different (e.g., photometric, dynamic, and microlensing) observations ranges from 0.5 to 2${\rm \times10^{10}\ M_{\odot}}$ \citep[][and see references therein]{licquia2015}. Interestingly, among these diverse measurements, the low mass end of $\sim$0.5${\rm \times10^{10}\ M_{\odot}}$ is all based on photometric data \citep{freudenreich1998,picaud2004,lopez-corredoira2007}. The idea behind the stellar mass estimate using photometric data as in previous works and spectroscopic data as in this work is essentially {the same}, star counting. This may explain why both approaches deliver a relatively low mass of the bulge. Comparing to the photometric data, spectroscopic observations are subject to a stronger selection effect but provide better distance that reduces foreground contamination. The selection effect can be mitigated by increasing the observed sample size and simplifying the target selection criteria. We anticipate that the ongoing and upcoming next generation of inner Galaxy spectroscopic surveys, such as SDSS-V and 4MOST, will significantly improve the estimate of the bulge stellar mass. 

The reason for the discrepancy in the disk stellar mass estimate is mainly two-fold. The main reason is that all previous estimates rely on the assumption of a single-exponential disk profile \citep{licquia2015,xiang2018}. As the inner Galaxy density profile within solar radius was not available before, an extrapolation from the outer disk exponential profile has been applied to estimate the total stellar mass. This extrapolation, however, could overestimate the disk stellar mass as the inner disk density profile may deviate from the outer disk profile with a lower density \citep{lian2024b}. 
By extrapolating the outer disk exponential profile using a scale length of { 2.1~kpc}, we obtain a total disk stellar mass of { 4.341${\rm \times10^{10}\ M_{\odot}}$, a factor of 2.2} higher than the estimate based on the reconstructed density profile. In addition, {} differences in the Galactic radial structure, including the local surface mass density, the scale length of the outer disk, and the Galactocentric radius of the Sun, may also contribute to the discrepancy in the disk stellar mass estimates. A higher local surface mass density was adopted in previous works (e.g., 34.6~${\rm M_{\odot}pc^{-2}}$ in \citet{licquia2015} and 35.7~${\rm M_{\odot}pc^{-2}}$ in \citet{xiang2018}). \citet{licquia2015} also assumed a { different} disk scale length of 2.15~kpc and a larger solar radius of 8.33~kpc, which further amplify the overestimation in disk stellar mass. 

A lower total stellar mass of the Milky Way was implied in some other Galactic properties that scales with stellar mass. The Milky Way hosts a supermassive black hole with a mass of $\sim4\ \times10^{6}\ {\rm M_{\odot}}$ \citep{gravity2023}. Assuming the previous stellar mass estimate of $\sim6\ \times10^{10}\ {\rm M_{\odot}}$, the Milky Way's central black hole is superficially small with a black hole to total stellar mass ratio of 6.7$\times10^{-5}$, a factor of 4 lower than the common ratio of 2.5$\times10^{-4}$ in low-redshift galaxies hosting Active Galactic Nuclei \citep{reines2015}. This difference becomes even more extreme when comparing to the inactive black holes in quiescent galaxies \citep{kormendy2013}. Adopting our revised total stellar mass estimate of $2.607\ \times10^{10}\ {\rm M_{\odot}}$ gives a new ratio of {1.6$\times10^{-4}$}, which becomes largely consistent with that of the local galaxies. Another implication of lower Galactic stellar mass is suggested by the integrated metallicity in both gas and stars of the Milky Way, which are found to be 0.2~dex lower than those of the local galaxies with a stellar mass of $6\ \times10^{10}\ {\rm M_{\odot}}$ (Lian et al. submitted). This difference is reduced to 0.15~dex in gas metallicity and 0.1~dex in stellar metallicity when using our revised estimate of the Galactic total stellar mass. Finally, {in terms of galaxy size,} the Milky Way also become more consistent with local galaxies with the lower stellar mass estimate \citep{lian2024b}. 

{ Given the observed circular velocity that determines the dynamical mass, the lower stellar mass suggests a larger fraction of dark matter in the inner Galaxy. Following \citet{huang2016}, we decompose the observed circular velocity curve within ${\rm R<17}$~kpc into two major components, the baryons and dark matter (DM). The former includes contributions from the bulge, stellar and gaseous disks. We further divide the stellar disk into low- and high-$\alpha$ components by adopting the chemical definition in \citet{lian2020b}. For the bulge, we adopt the 3D density profile and corresponding circular velocity from \citet{binney2011} (see equation~(12) in \citet{zhou2023}). The bulge stellar mass is assumed to be the value derived in this work. For the gas and stellar disk, we split the disk into a series of very thin slices with even step of 0.01~kpc and calculate their potential at height $z=0$~kpc based on a razor-thin disk assumption: 
\begin{equation}
d \Phi(R,z) = [\int_0^{\infty} I_kJ_0(kR)e^{-k|z|}dk]dz,
\end{equation}
where $J_0(kR)$ is Bessel function of zero order and $I_k=-2\pi G\int_0^{\infty}\rho(R,z)J_0(kR)RdR$. $\rho(R,z)$ denotes the 3D mass density as a function of radius and height. The calculation of the mass density distribution of the stellar disk is described in \textsection2.2, while that of the gas disk is calculated based on the structural parameters taken from \citet{binney2011}. The total potential is then integrated over all slices at different heights ($\Phi_{\rm total}(R,0) = 2\sum_i d\Phi(R,z_i)$). The corresponding circular velocity can be drived by the following 
\begin{equation}
    V_{\rm c,disk}^2(R) = R\frac{\partial \Phi_{\rm total}(R,0)}{\partial R}.
\end{equation}
For the dark matter component, we assume the canonical Navarro-Frenk-White (NFW) profile \citep{navarro1995} and the circular velocity can be described by 
\begin{equation}
    V_{\rm c,halo}^2 = \frac{4G\pi \rho_0 r_s^3}{r}[ln(1+\frac{r}{r_s})-\frac{r}{r_s(1+\frac{r}{r_s})}],
\end{equation}
where $\rho_s$ and $r_s$ are the characteristic DM density and scale radius, respectively. We the fit the observed circular velocity curve from \citet{zhou2023} with the two free parameters of $\rho_s$ and $r_s$ of DM density profile. The best-fitted parameters are presented in Figure~\ref{vc}. 
This best-fitted DM profile is different from previous works \citep[e.g.,][]{huang2016,zhou2023} with a higher characteristic DM density and shorter scale radius, suggesting a more concentrated DM profile with a larger contribution to the dynamic mass in the inner Galaxy. This is expected given the significantly lower stellar mass density of the inner disk estimated in this work. The local DM density is 0.011~${\rm M_{\odot}pc^{-3}}$, which is in agreement with previous estimates of ~0.01-0.014~${\rm M_{\odot}pc^{-3}}$ based on the local mass density measurements within 1~kpc \citep{read2014,mckee2015,bienayme2014,sivertssons2018}.  
}

\section{Summary}
In this work, we reconstruct the vertical mass density profile at solar radius and radial surface mass density profile across the Galaxy using unprecedented observations from Gaia and APOGEE surveys after {carefully correcting} for the selection function. The calculation of the selection function of the selected samples follows the procedure described in \citet{lian2022b}. Based on the reconstructed intrinsic mass density profiles, we revisit the local and global masses of living stars and stellar remnants in the Milky Way.

The local mass density of living stars and stellar remnants reveal a strong north-south asymmetry at height beyond 1~kpc. {This asymmetry can be explained either the geometric thick disk is significantly thicker in the north or offset from the thin disk.} 
Zooming into the profile close to the disk mid-plane, the data favor the exponential density model over the sech$^2$ model, {with the latter} underestimates the density on the disk plane. Integrating the density profile gives a precise estimate of the surface mass density of 31.563$\pm2.813$(syst.)$\pm$0.024(stoch.)~${\rm M_{\odot}pc^{-2}}$, of which 25.074 and 6.489~${\rm M_{\odot}pc^{-2}}$ are living stars and stellar remnants, respectively. These results are consistent with previous estimates with slightly {lower} values. 
 
Based on the radial surface mass density profile recalibrated by the local surface mass density {estimated with} Gaia data, we further derive the half-mass radius and global mass of the Milky Way. We obtain a half-mass radius of { 4.12$\pm$0.35(syst.)$\pm$0.15(stoch.)~kpc}, which is smaller than the half-light radius due to the more compact morphology of older populations. The obtained total stellar mass of the Milky Way is { 2.607$\pm$0.353(syst.)$\pm$0.148(stoch.)${\rm \times10^{10}M_{\odot}}$, of which 2.012 and 0.595~${\rm \times10^{10}M_{\odot}}$} are living stars and stellar remnants, respectively. Decomposing the mass density profile into disk and bulge components, their masses are { 1.935 and 0.673~${\rm \times10^{10}\ M_{\odot}}$}, respectively, implying a bulge-to-total ratio (B/T) of { 0.26}. Our estimate of the total stellar mass of the Galaxy is substantially lower than previous estimates of $\sim$6~${\rm \times10^{10}\ M_{\odot}}$. This is mainly because in previous works the inner Galaxy density profile was not available and assumed to follow the outer disk exponential profile. This extrapolation, however, may overestimate the disk mass by a factor of two, given that the inner disk profile was much flatter than the outer disk \citet{lian2024b}. This revised estimate of stellar mass {makes} the Milky Way a more typical galaxy in the local Universe in terms of many galaxy properties, including the central supermassive black hole mass, integrated gas and stellar metallicity, and galaxy size. { In addition, given the dynamic mass constrained by circular velocity, a lower stellar mass implies a larger fraction of dark matter in the inner Galaxy.} We expect that the ongoing and upcoming next generation of massive stellar spectroscopic surveys will further significantly improve the estimate of the stellar mass of the Milky Way, in particular {for} the bulge component. 

\section*{Acknowledgements} 
We are grateful to the anoymous referee for their constructive comments that have significantly improved the paper. J.L. acknowledges support by National Natural Science Foundation of China (No. 12473021), National Key R\&D Program of China (No. 2024YFA1611600), Yunnan Province Science and Technology Department Grant (No. 202105AE160021 and 202005AB160002), Key Laboratory of Survey Science of Yunnan Province (No. 202449CE340002), and the Start-up Fund of Yunnan University (No. CY22623101). 

Funding for the Sloan Digital Sky Survey IV has been provided by the Alfred P. Sloan Foundation, the U.S. Department of Energy Office of Science, and the Participating Institutions. SDSS-IV acknowledges support and resources from the Center for High-Performance Computing at the University of Utah. The SDSS web site is www.sdss.org.

SDSS-IV is managed by the Astrophysical Research Consortium for the 
Participating Institutions of the SDSS Collaboration including the 
Brazilian Participation Group, the Carnegie Institution for Science, 
Carnegie Mellon University, the Chilean Participation Group, the French Participation Group, Harvard-Smithsonian Center for Astrophysics, 
Instituto de Astrof\'isica de Canarias, The Johns Hopkins University, Kavli Institute for the Physics and Mathematics of the Universe (IPMU) / 
University of Tokyo, the Korean Participation Group, Lawrence Berkeley National Laboratory, 
Leibniz Institut f\"ur Astrophysik Potsdam (AIP),  
Max-Planck-Institut f\"ur Astronomie (MPIA Heidelberg), 
Max-Planck-Institut f\"ur Astrophysik (MPA Garching), 
Max-Planck-Institut f\"ur Extraterrestrische Physik (MPE), 
National Astronomical Observatories of China, New Mexico State University, 
New York University, University of Notre Dame, 
Observat\'ario Nacional / MCTI, The Ohio State University, 
Pennsylvania State University, Shanghai Astronomical Observatory, 
United Kingdom Participation Group,
Universidad Nacional Aut\'onoma de M\'exico, University of Arizona, 
University of Colorado Boulder, University of Oxford, University of Portsmouth, 
University of Utah, University of Virginia, University of Washington, University of Wisconsin, 
Vanderbilt University, and Yale University.

\bibliographystyle{aasjournal}
\bibliography{references_jl}{}
\end{document}